\documentclass{agujournal2018_DKoronczay_v2}
\usepackage{apacite}
\usepackage{aas_macros}
\usepackage{url} 
\usepackage{textcomp} 

\draftfalse

\journalname{Journal of Geophysical Research - Space Physics}

\begin{document}

\title{The source regions of whistlers}

\authors{ D\'avid Koronczay\affil{1,2}, J\'anos Lichtenberger\affil{1,2},
Mark A. Clilverd\affil{3},
Craig J. Rodger\affil{4},
Stefan I. Lotz\affil{5},
Dmitry V. Sannikov\affil{6},
Nina V. Cherneva\affil{6},
Tero Raita\affil{7},
Fabien Darrouzet\affil{8}, 
Robert C. Moore\affil{9}
}

\affiliation{1}{Department of Geophysics and Space Sciences, Eotvos University, Budapest, Hungary} 
\affiliation{2}{Research Centre for Astronomy and Earth Sciences, Hungarian Academy of Sciences, Sopron, Hungary}
\affiliation{3}{British Antarctic Survey (NERC), Cambridge, UK}
\affiliation{4}{Department of Physics, University of Otago, Dunedin, New Zealand}
\affiliation{5}{Space Science Directorate, South African National Space Agency, Hermanus, South Africa}
\affiliation{6}{Institute of Cosmophysical Research and Radio Wave Propagation, Far Eastern Branch of the Russian Academy of Sciences, Paratunka, Russia}
\affiliation{7}{Sodankyl\"a Geophysical Observatory, University of Oulu, Oulu, Finland}
\affiliation{8}{Belgian Institute for Space Aeronomy, Brussels, Belgium}

\affiliation{9}{Department of Electrical and Computer Engineering,
University of Florida, FL, USA}

\correspondingauthor{David Koronczay}{david.koronczay@ttk.elte.hu}

\begin{keypoints}
\item We present a method for identifying the geographical source regions of lightning generated whistlers.
\item Whistler source regions corresponding to 15 fixed whistler detector ground stations are determined.
\item Whistler transmission rates as a function of location are also determined.

\end{keypoints}

\begin{abstract}
We present a new method for identifying the source regions of lightning generated whistlers
observed at a fixed location.
In addition to the spatial distribution of causative lightning discharges,
we calculate the transmission ratio of lightning discharges into
ground detectable whistlers as a function of location.
Our method relies on the time of the whistlers and
the time and source location of spherics from a global lightning database.
We apply this method to whistlers
recorded at fifteen ground based stations in the AWDANet (Automatic Whistler
Detector and Analyzer Network) operating between 2007-2018
and to located lightning strokes from the WWLLN (World Wide Lightning Location Network) database.
We present the obtained maps of causative lightning and transmission ratios.
Our results show that the source region of whistlers
corresponding to each ground station
is around the magnetic conjugate point
of the respective station.
The size of the source region is typically less than 2000~km in radius
with a small fraction of sources extending to up to 3500~km.
The transmission ratio is maximal at the conjugate point
and decreases with increasing distance from it.
This conforms to the theory that
whistlers detected on the ground
propagated in a ducted mode through the plasmasphere
and thus the lightning strokes of their causative spherics
must cluster around the footprint of the ducts in the other hemisphere.
Our method applied resolves the whistler excitation region mystery that resulted from correlation-based analysis methods, concerning the source region of whistlers detected in Dunedin, New Zealand.
\end{abstract}

\section{Introduction}
Whistlers are very low frequency (VLF) electromagnetic waves originating from lightning discharges. 
These waves can penetrate into and through the ionosphere, entering the magnetosphere. They take a specific time-frequency shape through propagating in a dispersive medium, which is the magnetispheric plasma surrounding the Earth \citep{helliwell1965}.
Some of these waves may be trapped in field aligned density plasma irregularities, or ducts, extending between the two hemispheres. The trapping mechanism is explained by the theory of \citet{smith1960}. \citet{smith1961} lists evidence of the existence of ducts of enhanced density and describes whistler propagation in this structure.
The ducts guide the whistlers to the magnetic conjugate point, where, given the right conditions, they can re-enter through the ionosphere and become detectable on the ground in the conjugate hemisphere.
Another significant portion of the signals remain non-ducted.

Lightning generated whistlers are known to significantly affect the radiation belts through wave-particle interactions, causing acceleration \citep{trakhtengerts2003} and losses, be it oblique or magnetospherically reflected whistlers \citep{lauben2001,bortnik2006} or ducted whistlers \citep{helliwell1973,rodger2004}.
This whistler induced precipitation in turn influences the ionosphere \citep{helliwell1973,rodger2007}.
It has been shown that most lightning generated spherics leak into the ionosphere \citep{holzworth1999} and during strong lightning activity a significant fraction will reach the equatorial magnetosphere as whistlers \citep{zheng2016}.
They substantially affect the overall wave intensity in this region at the relevant frequencies \citep{zahlava2018}.
The shape of the whistler signals carry information on the magnetospheric plasma and has been used to investigate the plasmasphere \citep{helliwell1965,park1972,carpenter1988,lichtenberger2013}. Specifically, they serve as a ground based tool for mapping electron densities in the plasmasphere \citep{park1978}.
The detection and analysis of these waves have recently become a fully automated operation \citep{lichtenberger2008,lichtenberger2010}.
The fact that whistlers are used as a remote sensing  to study the plasmasphere and their role in controlling radiation belt populations provide a motivation to better understand their source and propagation.

The conditions for VLF signal propagation into the magnetosphere are better at high geomagnetic latitudes \citep{helliwell1965}, while lightning occurs predominantly in the tropics, at low geographic latitudes \citep{christian2003}. Thus, the resulting long term whistler rate at any location is a result of the two effects. 
Evidence suggests that there exists a low-latitude cutoff at geomagnetic latitude $16^{\circ}$ below which no whistlers can be observed on the ground, either due to a lack of appropriate
trapping and transmission conditions or a lack of ducts \citep{rao1974,helliwell1965}.
Furthermore, there are a number of variable conditions that can influence the transmission of whistlers to the ground, such as lightning activity at any given time, ionospheric conditions, the presence of ducts, etc.
The occurrence rate of whistlers is generally much
lower than the rate of conjugate lightning discharges.
Thus, it is of interest to know which factors determine
the reception of whistlers.
Since lightning remains a source effect, understanding its role is a necessary first if we are to untangle these various factors.

While the general picture of whistler propagation is understood, the exact location and extent of the source region is unknown.
A natural assumption is that the source is symmetrically centered on the conjugate point, but this has not been demonstrated. \citet{yoshino1976} found that lightning activity was displaced poleward and west from the conjugate point of Sugadaira, Japan, while \citet{oster2009} observed the source lightning distribution extending poleward and east from the conjugate point of Tihany, Hungary. \citet{gokani2015} observed a similar tendency for source lightning corresponding to very low latitude whistlers observed at Allahabad, India. While the propagation mechanism of very low latitude whistlers is not well understood, they showed the likely source region to be within 1000~km radius of the conjugate point.
Whistlers have been associated with lightning strokes occurring at 2000~km from the duct footprint by \citet{storey1953}, 2500~km by \citet{carpenter1989} and, in the case of whistlers at polar latitudes, several thousand km by \citet{allcock1960}.
Such propagation paths for whistler-mode waves were confirmed by \citet{clilverd1992}.
We note that even for whistlers that are eventually trapped in ducts, a part of the path in the ionosphere and magnetosphere may be non-ducted.
According to a review by \citet{holzworth1999}, upward going whistlers can be detected in the ionosphere at over 1000~km horizontal distance from the lightning sources, based on rocket experiments.
\citet{chum2006} manually identified 3500 fractional hop whistlers on the DEMETER satellite in low Earth orbit and paired them to lightning strokes from the EUCLID (European Cooperation for Lightning Detection) regional database,
finding that lightning discharges enter into the magnetosphere as whistler mode waves at distances up to 1500~km from the source.

An example of the holes in our understanding of whistler sources is the "Dunedin paradox". Lightning rate at the corresponding conjugate point is  orders of magnitude lower than the whistler rate observed in Dunedin, New Zealand. In addition, the time of the daily peak of the whistler rate at Dunedin is in disagreement with the peak of lightning activity at the conjugate point \citep{rodger2009}.
The results of \citet{collier2010} suggest tropical Mexico, over 7000~km from the conjugate point, as the source region, which, however, seem to contradict our understanding of the fundamental physics and also other observations \citep{morgan1956,antel2014}.

Apart from the direct matching of a small number of whistlers to lightning discharges, as done by \citet{storey1953,allcock1960,carpenter1989}, there have been few more general correlation studies.
\citet{ohta1990} found no correlation between whistler rates at Yamaoka, Japan, and lightning flash rates at the vicinity of its conjugate point in Australia. Whistler measurements were done between 1 to 15 January of every year from 1977 to 1987; while only monthly flash counts were available in the conjugate area, within a 50~km range, providing a limited dataset.
\citet{collier2006} compared the diurnal and seasonal rate of whistlers observed at a mid-latitude station (Tihany, Hungary, L=1.8) and the lightning activity in the assumed source region based on data from
WWLLN and LIS/TRMM (Lightning Imaging Sensor on the Tropical Rainfall Measuring Mission satellite), with results broadly consistent with expectations.

\citet{collier2009} presented a method of calculating Pearson correlation coefficients between whistler rates and lightning rates  separately for each of the grid cells spanning the globe.
They applied this method to whistlers recorded at Tihany station, Hungary, between 1 January 2003 and 19 May 2005.
The results showed significant positive correlation near the conjugate point of the station within a $\sim$1000~km radius, especially in the afternoon and early night. \citet{collier2010} applied the same method to whistler series from Dunedin, New Zealand, recorded between 20 May 2005 and 13 April 2009. In this case, the results showed a lack of correlation in the vicinity of the conjugate point (near Alaska Peninsula), weak correlation over the North Pacific, and strong correlation in Mexico.
Since the Pearson correlation coefficient is sensitive to large spikes that are often present in lightning discharge rates, \citet{collier2009,collier2010} used a boolean rounding, collating event counts within a predetermined time slot (${\Delta}t = 1 \textrm{min}$) to 0 or 1 to overcome this problem, at the cost of losing some amount of information. In a subsequent study by \citet{collier2011} of whistlers recorded at Rothera, Antarctica, between
13 May 2008 and 30 December 2009, no such reduction was applied
to the event rates. In this case, significant positive correlation
over the Gulf stream, an active lightning center near the conjugate	 point of Rothera, was observed.
In all of the studies mentioned \citep{collier2009,collier2010,collier2011}, the method lead to other, additional areas of positive correlation, far away from the conjugate points, that may or may not be actual sources.
A similar study by \citet{vodinchar2014} analysed whistlers detected at Karymshina, Kamchatka, between 1 to 11 March 2013 and 1 to 30 September, 2013. In addition to Pearson correlation coefficients with boolean rounding, their analysis applied Spearman rank correlation coefficients to the real (unrounded) whistler time series and lightning time series calculated for each continent. No significant positive correlation was found near the conjugate point of the station, or in surrounding Australia in general.

\citet{collier2011} also present another, different method for mapping causative lightning strokes. Instead of relying on correlations, this direct method simply registers every lightning stroke that falls within a predetermined time window preceding each whistler recorded at a specific ground station, Rothera, in this case, and the geographic distribution of those lightning strokes are presented on a density map. The results show a strong reminiscence of the lightning density distribution around the Gulf Stream, an area with high frequency of lightning, leading to a much more well defined result than those of the correlation-based methods listed above.

The goal of our paper is a better understanding of the positions of causative lightning strokes that lead to whistlers.
Once the source regions are reliably identified, subsequent studies can look into correlations
between whistler counts and lightning in the source region.
With the availability of long-term global lightning data through WWLLN, and our large dataset of AWDANet
whistler measurements from fifteen stations around the world recorded over twelve years, we can extend our investigation into a significantly larger scale than previous studies.

\section{Data}

The WWLLN (World Wide Lightning Location Network, http://www.wwlln.com/) is a global network consisting of VLF sensors. The network uses the time of group arrival method from at least five stations to locate individual lightning strokes. Due to the low attenuation of VLF waves in the Earth-ionosphere waveguide, it has global sensitivity, as opposed to regional lightning detector networks often operating at higher frequencies.
A temporary drop-out of any single station has only a slight effect on the detection efficiency \citep{wwlln4}. Therefore, the lightning stroke time series used in this study was considered continuous, without any data gaps.

The number of stations in the network steadily increased from 23 stations in 2005 to $\sim$70 stations at present. After the upgrades in the processing algorithm (used to reprocess the entire raw dataset), this expansion resulted in the total number of located lightning strokes of 36~million in the year 2005 increasing to 208~million by 2017.
WWLLN is capable of detecting both cloud-to-ground (CG) and intracloud (IC) flashes of sufficient strength. The total detection efficiency (CG+IC)
is estimated to have increased from 2.6\% in 2005 to about 15\% in 2017 \citep{wwlln1,wwlln2,wwlln3,wwlln5}.
In addition, detection efficiency is strongly dependent on peak current, and can be as high as 35\% for currents exceeding $-130 \textrm{kA}$ \citep{wwlln3}. Detection efficiency in the long term is lowest over ice covered surfaces such as Greenland and Antarctica \citep{wwlln4}.
Location accuracy is estimated at $<10 \textrm{km}$,
much smaller than the pixel sizes on Figures
\ref{fig:method}-\ref{fig:regional_hu}.
In the present study we used $2 \times 10^{9}$ lightning strokes with locations from WWLLN, recorded between 2007 and 2018.

\begin{table}[hbt]
\caption{Whistler recordings processed in this study, in regional grouping}

\begin{tabular}{ l l l l r r}
\hline
 Station & Geodetic    & L-value$^{a}$  & Years     & Total number & Max. transmission\\
         & coordinates &                & processed & of whistlers & ratio [\%] \\
\hline
 Dunedin & 45.7\textdegree S, 170.5\textdegree E & 2.78 & 2007-2017 & 3,660,000 & 75 \\ 
 \hline
 Karymshina & 53.0\textdegree N, 158.7\textdegree E & 2.18 & 2012-2016 & 3,110,000 & 25 \\ 
 \hline
 Palmer & 64.8\textdegree S, 64.0\textdegree W  & 2.52 & 2009-2010 & 17,600,000 & 50 \\ 
 Rothera & 67.5\textdegree S, 68.1\textdegree W & 2.82 & 2008-2016 & 43,300,000 & 20 \\ 
 Halley & 75.6\textdegree S, 26.6\textdegree W & 4.75 & 2012-2015 & 4,300,000 & 30 \\ 
 SANAE & 71.7\textdegree S, 2.8\textdegree W & 4.60 & 2006-2016 & 1,780,000 & 20 \\ 
 \hline
 Sutherland & 32.4\textdegree S, 20.6\textdegree E & 1.78 & 2007-2011 & 30,000  & 0.5 \\ 
 Grahamstown & 33.3\textdegree S, 26.5\textdegree E & 1.82 & 2015-2018 & 124,000  & 0.5 \\ 
 Marion Island & 46.9\textdegree S, 37.9\textdegree E & 2.68 & 2009-2016 & 3,540,000 & 12 \\ 
\hline
 Tihany & 46.9\textdegree N, 16.9\textdegree E & 1.83 & 2007-2017 & 820,000 & 4\\ 
 Gyergy\'o\'ujfalu & 46.7\textdegree N, 25.5\textdegree E & 1.84 & 2007-2016 & 120,000 & 2\\ 
 Nagycenk/Muck & 47.6\textdegree N, 17.7\textdegree E & 1.81 & 2007-2018 & 285,000 & 3 \\ 
 Humain & 50.2\textdegree N, 5.2\textdegree E & 2.09 & 2011-2018 & 128,000 & 4 \\ 
 Eskdalemuir & 55.3\textdegree N, 3.2\textdegree W & 2.72 & 2011-2018 & 10,000 & $\geq 12$\\ 
 Tv\"arminne & 59.8\textdegree N, 23.0\textdegree E & 3.32 & 2013-2018 & 346,000 & $\geq 15$\\ 
\hline 
\multicolumn{6}{l}{
$^{a}$At 100~km altitude and at the epoch of 2015
using the IGRF-12 geomagnetic model \citep{IGRF12}.}
\end{tabular}
\label{table:awdanetdata}
\end{table}

The AWDANet (Automatic Whistler Detector and Analyzer Network)
is a global ground-based network of VLF stations that automatically detect and analyze whistlers \citep{lichtenberger2008}. In the first data processing segment, the detection of whistlers yields a time series of whistler traces. The second segment, the analysis involves the scaling and inversion of each whistler signal, and yields plasmaspheric electron densities along the propagation field line, the L-value of the field line, and an estimation of the time of the originating lightning stroke \citep{lichtenberger2010}. While the latter can be of help when associating causative strokes to whistlers, currently only about 1 to 5\% of the input is successfully inverted by the algorithm, significantly reducing the statistics. Therefore, we chose to use the results of only the detector segment, the time series of whistler traces.
We note that both networks use GPS timing, and the time accuracy of AWDANet whistlers is limited by the spread of the whistler traces to $ \sim 1 \textrm{ ms}$, while the accuracy of the reconstructed time of lightning strokes in WWLLN is $ \ll 1 \textrm{ ms}$.

AWDANet has been in real-time operation providing prompt results since 2014.
To extend this dataset, we have also processed available raw data between 2007 and 2014.
The earlier (2002-2007) measurements were not based on GPS precision timing and therefore we excluded those from the analysis.
Nevertheless, our method should in principle work with less accurate times, too. 
Table \ref{table:awdanetdata} lists the detector stations in regional groupings, the years of observations processed, and the total number of whistlers. 
Altogether we used 77~million whistler traces in this study from the fifteen stations combined.
Finally, the table also shows lightning to whistler transmission rates for each station, a result of the calculations presented in the following section.

In addition to AWDANet stations, data from Palmer Station, Antarctica are also included in this report. The VLF system at Palmer Station consists of two orthogonal IGY loop antennas with a sensitivity of $5.7 \times 10^{-18} \textrm{T} \textrm{Hz}^{-1/2}$ at 10~kHz.  The frequency response of the Palmer VLF system is flat between 130~Hz and 45~kHz, and data are recorded continuously at 100~kHz with 16-bit precision. Timing is supplied by a GPS-trained oscillator with $10^{-12}$ frequency precision.

\section{Method}
\label{section_method}

\begin{figure}[htb!]
    
    \includegraphics[width=9.5cm,height=7cm]{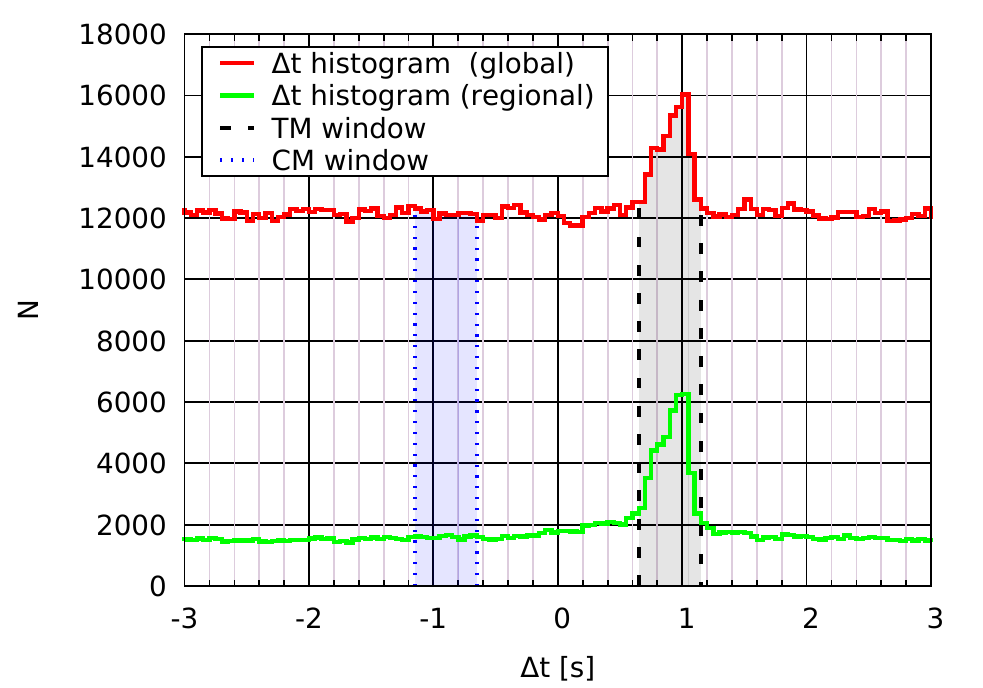}

\caption{Histogram of time differences between whistlers and lightning strokes, using WWLLN global data (red) and data restricted to the conjugate region (green). N is the number of lightning-whistler pairs. The peak is due to the tendency of whistlers to occur after causal lightning, while the background is caused by chance matches between unrelated whistlers and lightning. The peak is much more prominent on the regional map. Black dashed lines represent a time window for the selection of source strokes (TM, total matches). Blue dotted lines show a window of identical length (CM, chance matches) but with whistlers preceding lightning strokes to exclude any causality, used for the statistical removal of chance matches.}.
    \label{fig:dthist}
\end{figure}

Correlation-based methods described in the Introduction have a number of weaknesses. First, the resulting correlation maps sometimes include areas of negative correlation that have no relevant physical meaning. Second, more importantly, areas of positive correlation do not necessarily imply causality. For example, it is conceivable that the diurnal variation of the lightning flash rate at one location is, simply by chance, similar to the diurnal variation of the whistler rate at another location, the latter arising from a combination of source lightning flash rate, diurnal ionospheric changes and other propagation effects in the ionosphere and the plasmasphere, leading to some level of positive correlation between the two. Thirdly, by the same logic, any real positive correlation between whistlers and causative lightning strokes may be damped by time dependent propagation effects between the source and the detection.

For this reason, instead of relying on perceived correlations, we focus on the direct method, explained in \citet{collier2011} as a second method. This procedure attempts to directly pair up whistlers with their corresponding source lightning based solely on their timing. In an additional step we statistically correct for false matches, as explained in the next section.

\subsection{Mapping causative lightning strokes}
\label{subsection_causativemap}

\begin{figure}[htb!]
    
    \includegraphics[trim=2cm 0 -2cm 0 ,width=19cm,height=11.3cm]{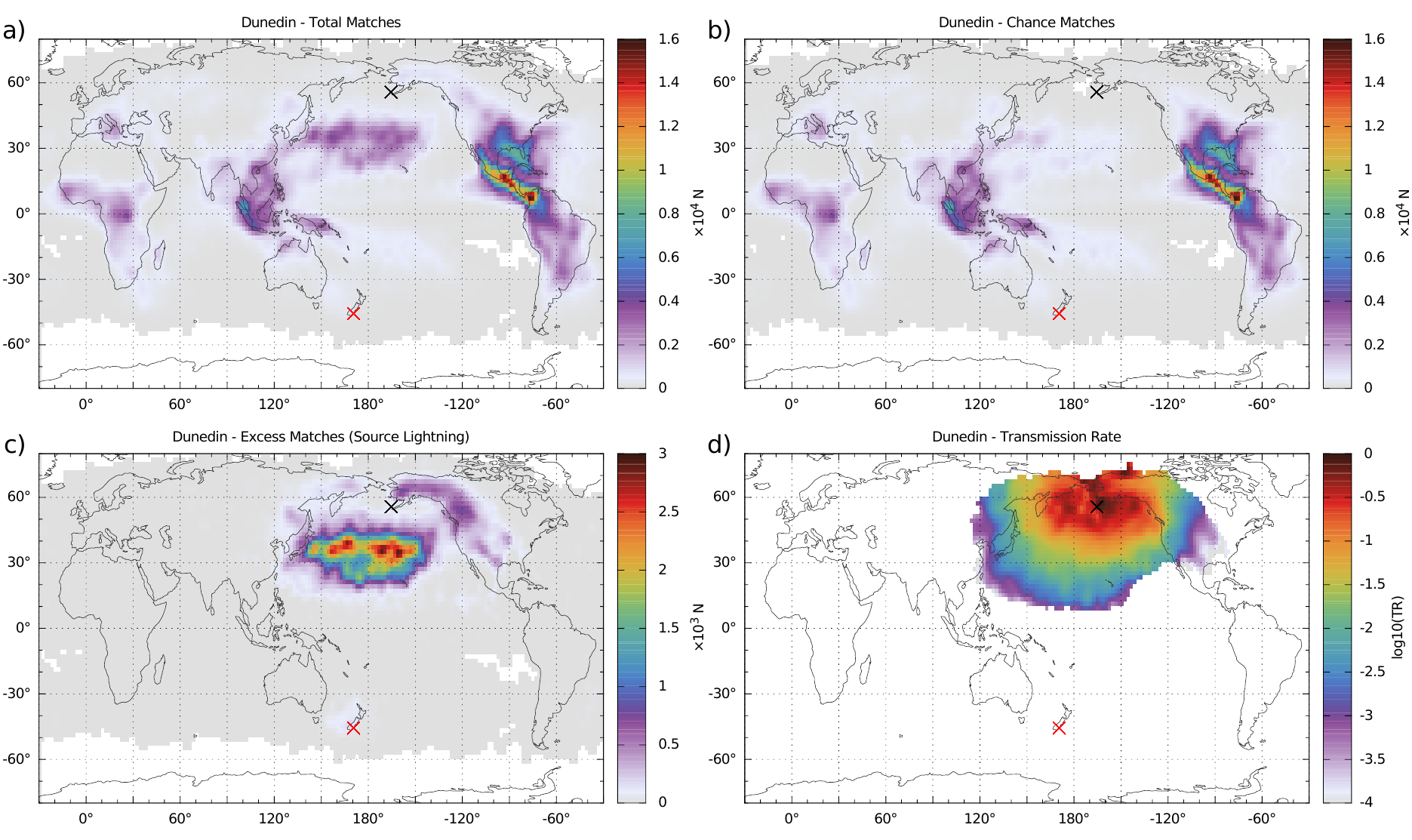}
    
\caption{Demonstration of the steps of our method. (a) Total matches (TM): distribution of all lightning strokes in the positive time window (lightning preceding whistlers, see Figure \ref{fig:dthist}) Pixel size is $2^{\circ}\times2^{\circ}$, color represents the number of matched strokes (N) in the given pixel. (b) Chance matches (CM): distribution of lightning activity in the negative time window (lightning following whistlers, excluding causality, representing purely chance matches between the two). (c) Excess matches (EM): difference between TM and CM. (d) Transmission rate (TR), or the number of excess strokes divided by the climatology shown in Figure \ref{fig:method_part2}, or the total number of WWLLN lightning strokes over the same time period.
Transmission rate (R) is shown only in the area of significant source lightning.
All maps are smoothed using a $3\times3$ pixel Gaussian kernel.
The red cross marks the location of the whistler recording station, the black cross marks its geomagnetic conjugate point using the IGRF-12 geomagnetic field model \citep{IGRF12}.}
    \label{fig:method}
\end{figure}

\begin{figure}[htb!]
    
    \includegraphics[width=9.5cm,height=5.65cm]{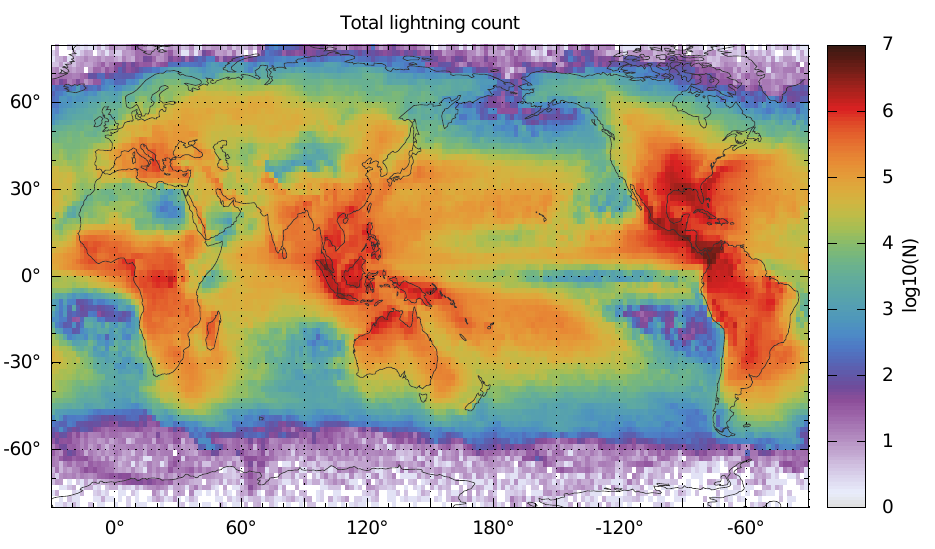}
    
\caption{"Climatology", or total number of WWLLN lightning strokes over the entire measurement period of the whistler station in question (Dunedin, 2007-2017). Periods when the whistler recording station was offline were excluded. Such maps are used for the normalization of lightning source maps to produce transmission rate maps, an example of which is shown in Figure \ref{fig:method}d. Pixel sizes are same as for Figures \ref{fig:method}a-\ref{fig:method}d, $2^{\circ}\times2^{\circ}$.
Pixel colours represent absolute number of located lightning discharges within the pixel, without correction for pixel sizes varying with latitude. Apart from this latitudinal factor, this map is similar to lightning density climatologies normalized to flashes $\textrm{km}^{-2}$~ $\textrm{yr}^{-1}$, see e.g. the lightning climatology obtained from WWLLN on Figure 2b in \citet{wwlln_climatology}.}
    \label{fig:method_part2}
\end{figure}

The only adjustable parameter in the direct technique is the
time window prior to each whistler in which lightning strokes are considered. This should be different for every station and is thus set separately for each one. In theory, a practical time window could be calculated as the minimum and maximum travel times of the source spheric travelling along the field line, depending on possible values of plasmaspheric density and field line L-value. Since there are no sharp limits on these values, we instead experimentally determine the appropriate time window from the data. 
Given a list of $n$ whistler times and $m$ lightning times, first an $n \times m$  matrix is computed of every possible time difference between whistlers and lightning strokes, such that the value of matrix element $M_{ij}$ is the $t_{\textrm{whistler}i} - t_{\textrm{lightning}j}$ difference of the time of the \textit{i}-th whistler and the \textit{j}-th lightning.
Next, an occurrence histogram of the time differences in the matrix is computed.
(A similar construct was used earlier by \citet{chum2006}.)

Figure \ref{fig:dthist} shows such a histogram, with a time bin of 50~ms, limited to practical values of a couple of seconds around zero time difference.
If the two time series were independent,
the time differences would be distributed randomly and uniformly on the histogram. The peak in the distribution is due to the fact that whistlers tend to occur a short time after their causative spheric, with the peak location corresponding to the most common travel time from the source through the magnetosphere to the location where it is recorded by a ground station.
Based on the location and width of the peak an appropriate window of a few hundred millisecond was determined for each whistler detector station separately.

Next, we create a series of density maps.
The first map represents the matched lightning strokes. For each grid cell on a geographic map, the total number of WWLLN strokes occurring within the predetermined time window preceding a whistler is calculated and the value is assigned to the cell. The obtained map of total matches, (TM, see Figure \ref{fig:method}.a for the TM map for Dunedin) should contain the actual source lightning strokes, but also a lot of chance coincidences, especially in regions of intense lightning activity. In order to remove these chance coincidences from the map, we try to estimate their expected value. Naturally, at locations of high lightning activity, such as the three main tropical lightning centers (visible on Figure \ref{fig:method_part2} representing long-term lightning activity), the probability of chance coincidences is higher, explaining some of the patches on Figure \ref{fig:method}.a.
A static map of lightning activity, however, cannot be directly used for the estimation of chance matches, since both whistler and lightning rates exhibit strong diurnal and seasonal variations, necessitating a combination appropriately weighted seasonal and hourly climatologies.
Instead of such a complicated procedure, we simply calculate matches between lightning strokes and whistlers in another time window, representing the "background noise". By choosing a window (see Figure \ref{fig:dthist}) that corresponds to negative time
difference between whistlers and lightning strokes,
we can be sure that no causative strokes are included in the selection. For each event, this window for background noise calculation precedes the other window
by merely a few seconds, which ensures that the global lightning activity and thus the likelihood of chance matches do not deviate significantly from that within the other (causative) window.
Detection efficiency of whistlers and spherics should also remain constant over such short period of time. Figure \ref{fig:method}.b shows the density map of chance matches (CM) calculated this way, again for the Dunedin station.

Finally, we subtract the two density maps to obtain a map representing only the excess matches (EM): $\textrm{EM}$~$=$~$\textrm{TM}$~$-$~$\textrm{CM}$.
Since these excess matches are due to the causative lightning strokes, the obtained map represents the geographic distribution of the causative lightning strokes corresponding to the whistlers detected at the given ground station. We term these "source lightning".

Remarkably, in the above procedure, there was no input specifying where the whistler time series was recorded, we simply used the Dunedin whistler time series as observed. Nevertheless, the resulting map on Figure \ref{fig:method}c clearly shows the source lightning distribution surrounding the conjugate point of Dunedin, New Zealand. This correspondence repeats when we undertake these calculations for each of the fifteen AWDANet stations, a strong validation of our method. Note that the colour scale in Figure \ref{fig:method}c is $\sim$5 times smaller than that of Figures \ref{fig:method}a and \ref{fig:method}b.
This emphasises the large number of chance matches, and also demonstrates how the chance matches can so easily leading to misleading results when simple correlation approaches are used.

\subsection{Mapping whistler transmission rates}

The procedure described so far tells us where the lightning source regions corresponding to a given whistler recording station are located. We can also calculate the geographic distribution of an additional parameter, the lightning-to-whistler transmission ratio. The low absolute number of source lightning in the vicinity of the conjugate point of Dunedin, for example, is not in itself surprising given the fact that the region shows very limited lightning activity in general.
This motivates a normalization of the source lightning counts with the total population of lightning strokes.
Figure \ref{fig:method_part2} shows an example of the latter, a "climatology" (CL) calculated as the total number of WWLLN lightning strokes over the period under study when the AWDANet station in question was in operation. Periods of data gaps in the whistler data (due to instrumental problems, data loss, etc.) are excluded from the count. This is the lightning stroke population which could theoretically generate the whistlers present in the time series.
Note that all of the maps on Figures \ref{fig:method}, \ref{fig:method_part2} and \ref{fig:globalmap} were calculated using $2^{\circ}\times2^{\circ}$ geographic grid cells.

Dividing the source lightning count (obtained by the procedure described in Subsection \ref{subsection_causativemap}) by the total lightning population over the same time period, we obtain a so-called transmission ratio, or the percentage of lightning strokes that generated whistlers observable at the relevant station. 
By doing so for each grid cell of the two maps (excess matches and climatology), this transmission rate (TR) can also be represented on a density map: $\textrm{TR}$~$=$~$\textrm{EM}$~$/$~$\textrm{CL}$. Figure \ref{fig:method}.d shows an example of the results obtained by this procedure, again for the Dunedin AWDANet station.
The transmission rate is calculated only for pixels with significant source lightning levels.

To a first approximation, the obtained TR is insensitive to the varying detection efficiency of the WWLLN network, since both the source lightning and the total lightning counts should be affected similarly.
However, it is possible that the sub-population of lightning strokes capable of whistler generation have different detection efficiency.
WWLLN detection efficiency depends on the lightning peak current, for example, which may affect whether an observable whistler is generated.

The whistler detection efficiency of AWDANet is also difficult to ascertain. Our preliminary study \citep{lichtenberger2008} showed that the whistler detector trace finder at the Tihany station works at $ < 3\%$ false positive and $ < 5\%$ false negative detection rate.
Nevertheless, unlike with lightning detection networks, we cannot compare our measurements to independent datasets. The aforementioned detection rate estimates are based on the assumption that whistler traces on the spectrograms recognized by the human eye constitute the total population of whistlers. Clearly, this is not necessarily true, as some traces, especially those having smaller amplitudes, may be swamped by noise in the same frequency band. How many traces are lost to the noise is not known. In our experience, the local electromagnetic noise levels of both artificial and natural origin are different at each station, and can be strongly time dependent. 

Thus, the level of completeness of both time series is difficult to ascertain. Therefore, the obtained TR should be taken as relative values, not necessarily directly comparable between stations.

\section{Results}
\label{section_results}

\begin{figure}[p]
    
     \includegraphics[trim=2cm 0 -2cm 0,width=19cm,height=22.6cm]{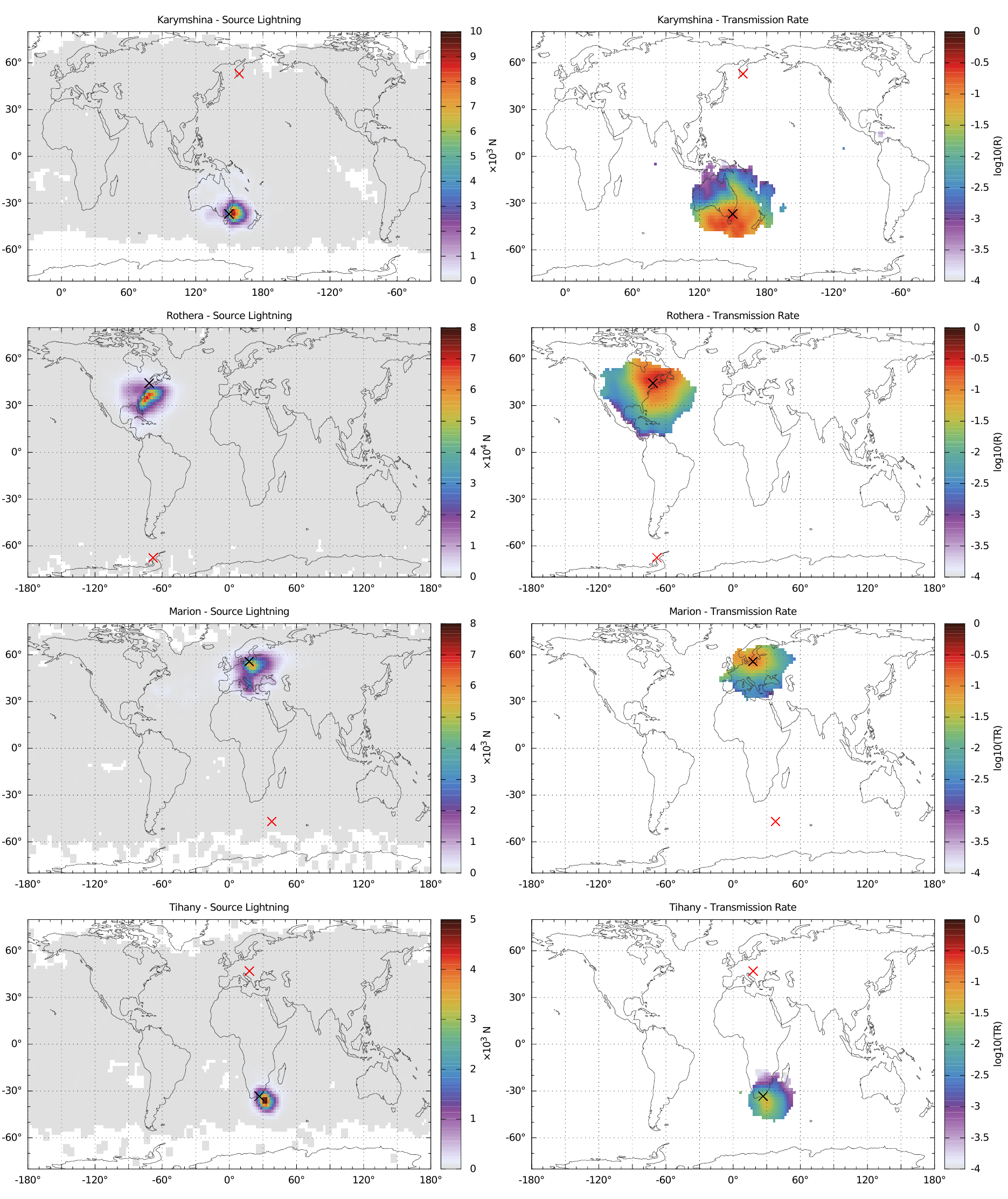}
\caption{Distribution of source lightning (EM, left) and transmission rate (TR, right) on a global map, in four regions. The left hand panels are in the same format as Figure \ref{fig:method}c, while the right hand panels are in the same format as Figure \ref{fig:method}d.}
     \label{fig:globalmap}
    
\end{figure}

The procedure described in Section \ref{section_method} is repeated for each station.
Figure \ref{fig:globalmap} shows the source lightning distribution and transmission rates for four stations, located in four different regions across the globe. These (and Figures \ref{fig:method}c-\ref{fig:method}d) demonstrate that the lightning which are whistler sources lie within a few thousand kilometers of the geomagnetic conjugate point, as expected. No sources or discernible transmission rate can be observed outside this area. The global lightning centers in the tropics play little to no role in the generation of ground detectable whistlers, at least for these middle to lower middle latitude stations, due to their large distance from these latitudes.

\begin{figure}[!htb]
 \includegraphics[trim=2cm 0 -2cm 0,width=19cm,height=11.3cm]{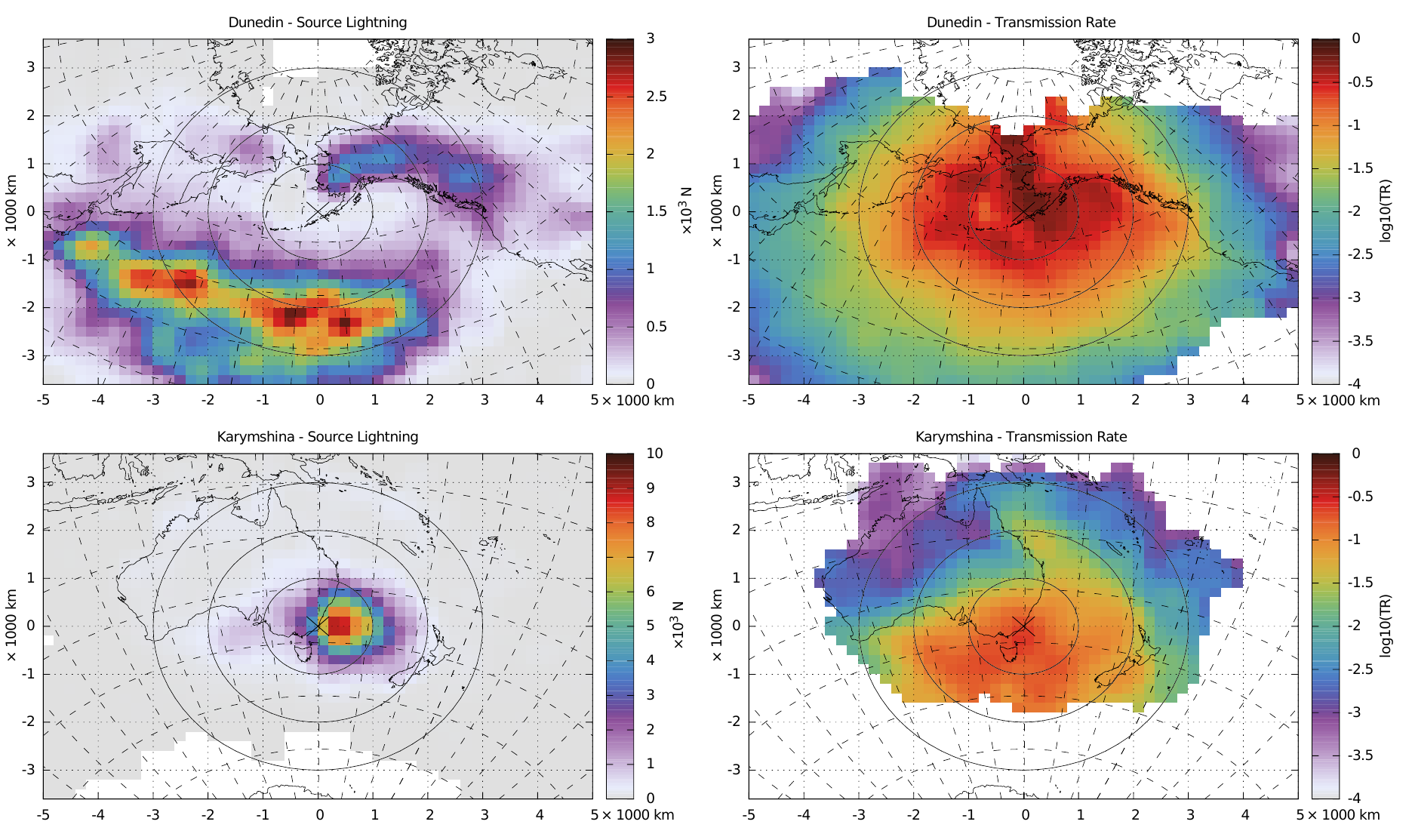}
\caption{(top) Regional distribution of source lightning (EM) and transmission rate (TR) for whistlers detected at Dunedin, New Zealand (showing its conjugate region near Alaska). (bottom) Same for Karymshina, Kamchatka (showing its conjugate region near Australia).}
 \label{fig:regional_westpacific}
\end{figure}

\begin{figure}[htb]
    
    \includegraphics[trim=2cm 0 -2cm 0,width=19cm,height=22.6cm]{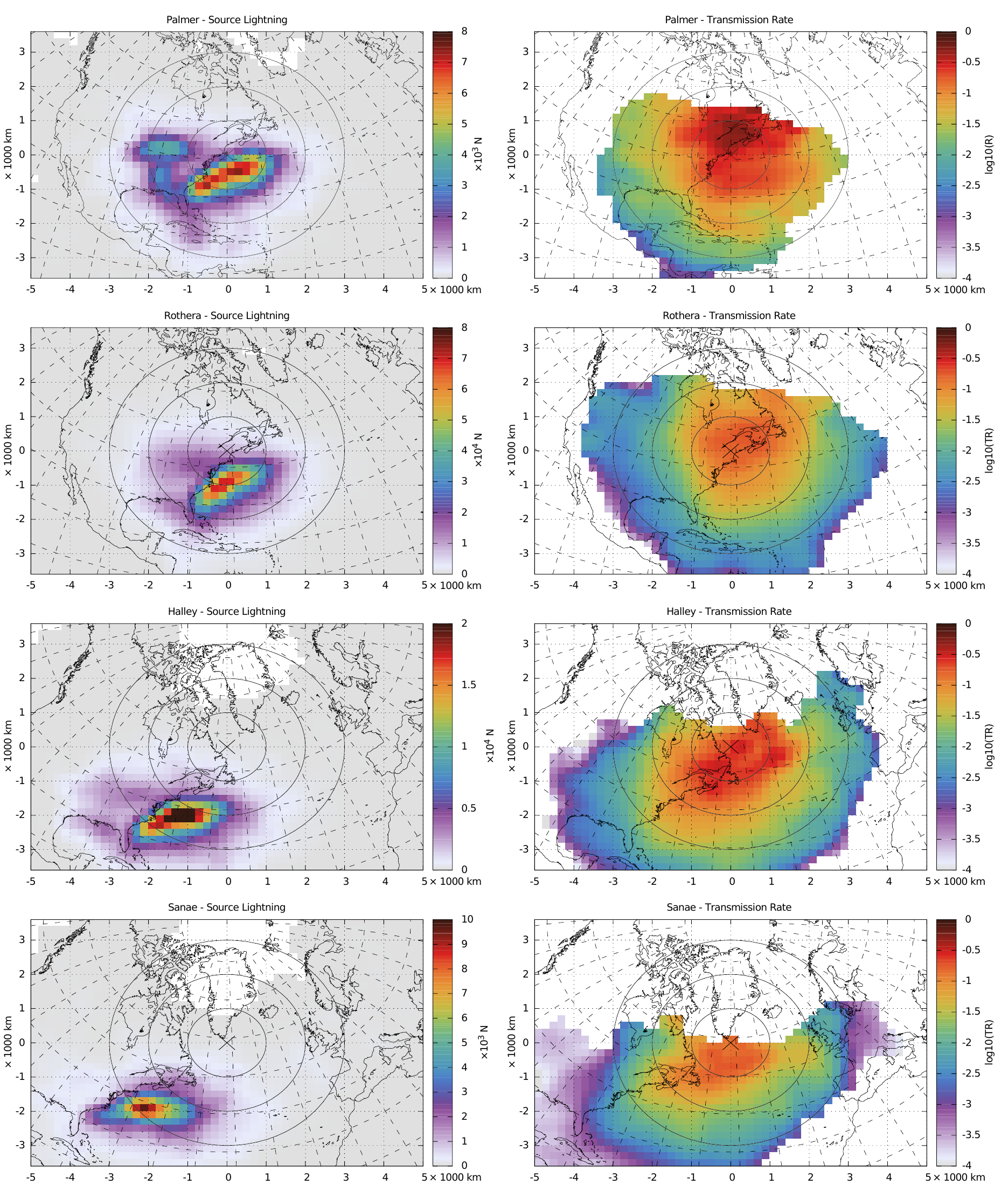}
    
\caption{Regional distribution of source lightning and transmission rate for whistlers detected at stations in West Antarctica}
    \label{fig:regional_americas}
\end{figure}

\begin{figure}[htb]
    
    \includegraphics[trim=2cm 0 -2cm 0,width=19cm,height=16.95cm]{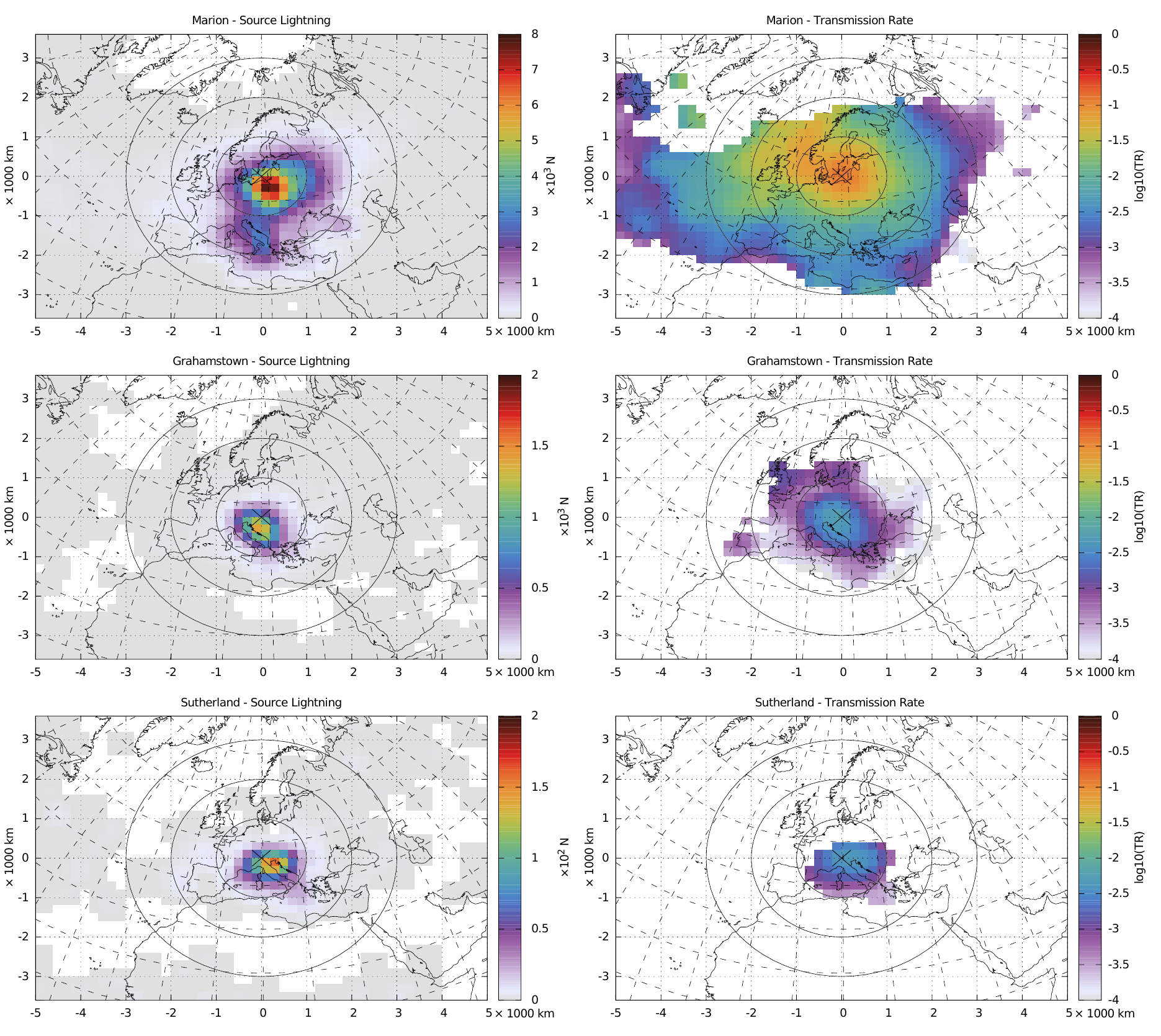}
    
\caption{Regional distribution of source lightning and transmission rate for whistlers detected in southern Africa.}
    \label{fig:regional_southernafrica}
\end{figure}

\begin{figure}[htb]
    
    \includegraphics[trim=2cm 0 -2cm 0,width=19cm,height=16.95cm]{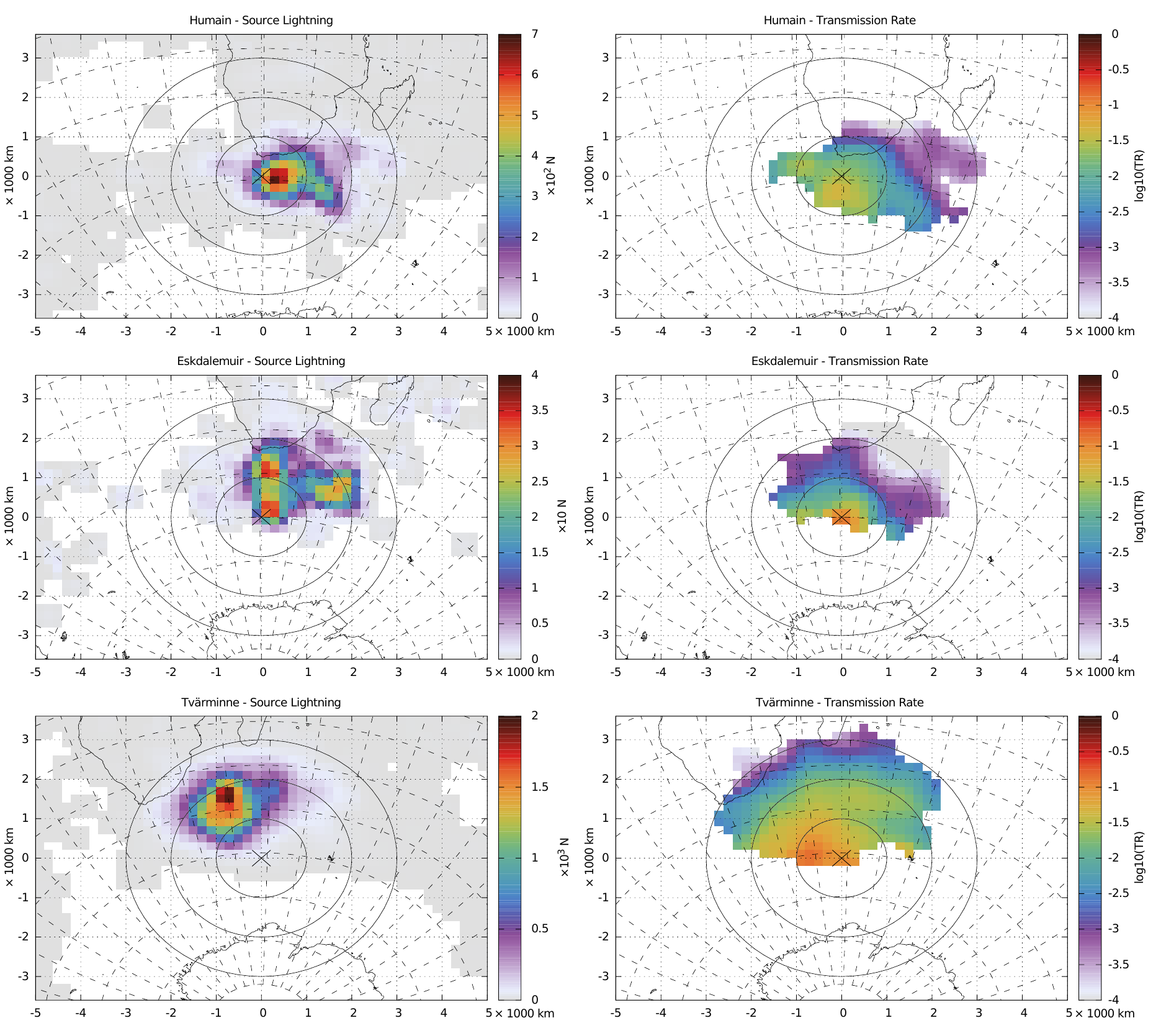}
    
\caption{Regional distribution of source lightning and transmission rate for whistlers detected at European stations.}
    \label{fig:regional_europe}
\end{figure}

\begin{figure}[htb]
    
    \includegraphics[trim=2cm 0 -2cm 0,width=19cm,height=16.95cm]{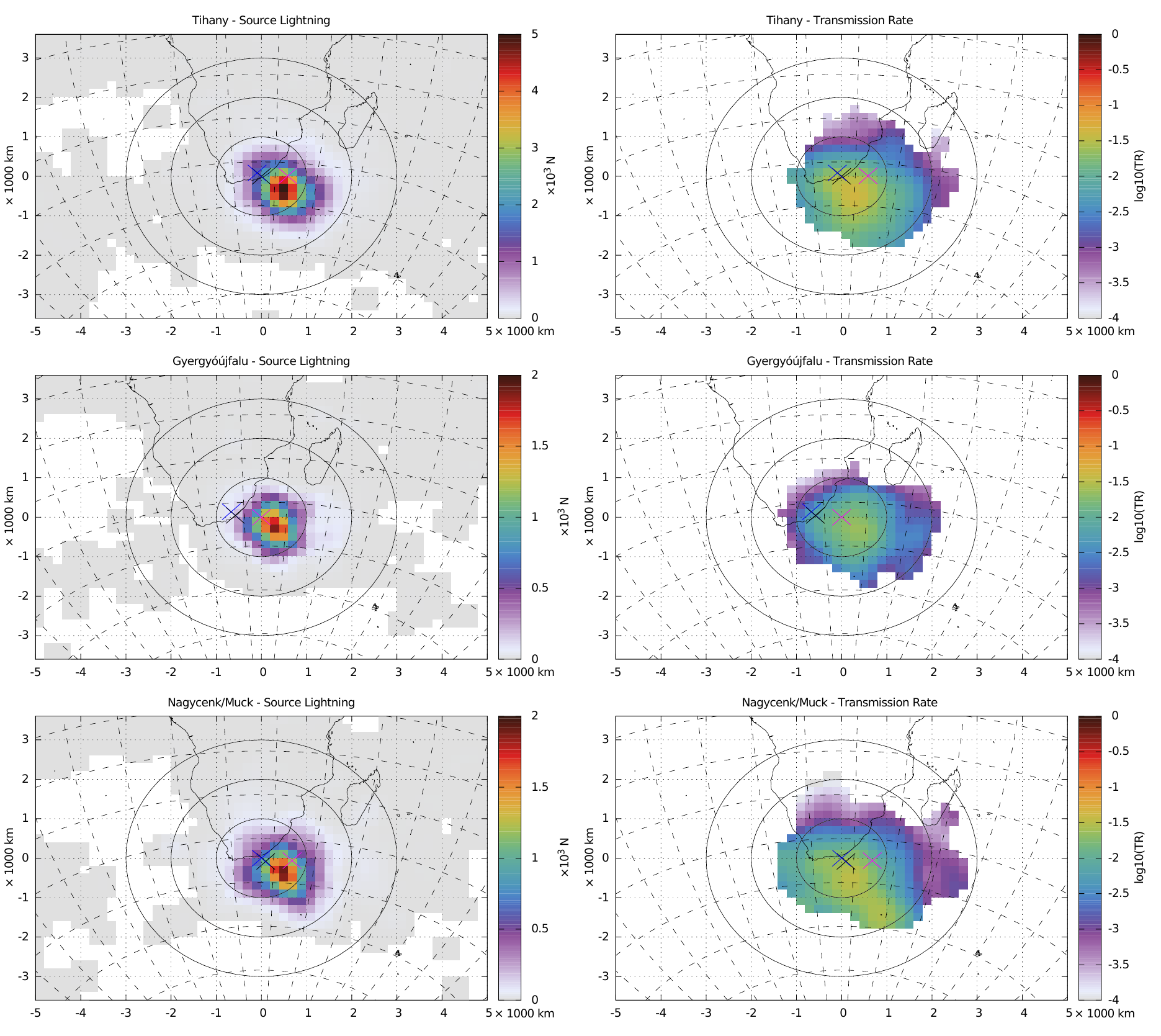}
    
\caption{Regional distribution of source lightning and transmission rate whistlers detected at the European stations of Tihany (black cross), Nagycenk/Muck (blue cross) and Gyergy\'o\'ujfalu (magenta cross).}
    \label{fig:regional_hu}
\end{figure}

Having established a global picture, the following maps are limited to the conjugate region for better viewing. Figures \ref{fig:regional_westpacific} to \ref{fig:regional_hu} are in azimuthal equidistant projection, which keeps true great circle distance and azimuth from the center point, the conjugate point of the relevant whistler detector station. Also, this projection has little areal distortion within a few thousand kilometers of the center point. We used $200\times200$~km pixels and applied a $3\times3$ Gaussian smoothing kernel to each map. Three concentric circles are plotted around the conjugate points, representing distances of 1000, 2000 and 3000~km. For better comparison, we used the same coordinate ranges and in the case of the transmission maps the same logarithmic scale for each map.

The maps are presented in regional groups. The top panels of Figure \ref{fig:regional_westpacific} show the source lightning distribution (EM) and transmission rate (TR) of whistlers detected at Dunedin, New Zealand, at its conjugate area located near Alaska. The bottom panels of Figure \ref{fig:regional_westpacific} are the same for whistlers detected at Karymshina, Kamchatka, with the conjugate area being located near Australia. Figure \ref{fig:regional_americas} shows conjugate areas in North America corresponding to stations in West Antarctica: Palmer, Rothera, Halley and SANAE. Figure \ref{fig:regional_southernafrica} shows conjugate areas in Europe corresponding to stations in Southern Africa: Sutherland, Grahamstown and Marion Islands. Figure \ref{fig:regional_europe} shows conjugate areas in Southern Africa corresponding to the European stations of Humain, Eskdalemuir and Tv\"arminne. Finally, Figure \ref{fig:regional_hu} shows the source lightning and transmission rate distribution for further three, closely separated stations in Central Europe: Nagycenk/Muck, Tihany and Gyergy\'o\'ujfalu.

\section{Discussion}

The regional maps show that the source lightning distribution is dominated by the nearest conjugate region of high lightning activity.
This is in agreement with the similar conclusion of \citet{collier2011} based on the second method listed there. Our method, however, does not produce any stray regions of whistler sources away from the conjugate region, unlike the first method of \citet{collier2011}. In some cases, the majority of the source lightning can be significantly offset from the actual conjugate point, such as in the case of Rothera, Halley and SANAE (where lightning above the Gulf current dominates), or Dunedin (where patches in the North Pacific Ocean dominate). The transmission maps, however, do not show such offsets in the distribution, and instead the transmission rate decreases largely monotonously with increasing distance from the conjugate point. The transmission rate is largest at or near (within 1000 km) of the conjugate point. No significant poleward offset can be observed in the distribution of the transmission rate. Nevertheless, such an offset cannot be entirely excluded, since in many cases, the poleward parts of the distribution are missing due to very low lightning activity at high latitudes. Also, the AWDANet whistler detector was optimized for whistlers with $ L < 4.5 $ and thus some part of the population may be missing.

The maps on the top panel of Figure \ref{fig:regional_westpacific} seem to partially answer the "Dunedin paradox" \citep{rodger2009}. The immediate area of the conjugate point of Dunedin indeed exhibits very low lightning activity. Nevertheless, as the transmission rate map shows, a large number of lightning discharges from further away will find their way to Dunedin, albeit with decreasing efficiency as we get further from the conjugate point.

In some cases (Palmer, Rothera, Halley, SANAE, see Figure \ref{fig:regional_americas}, and Dunedin, see Figure \ref{fig:regional_westpacific}), significant levels of transmission rate extend over 3000~km. On the other hand, the maximal transmission rate, and possibly as a consequence, the geographical extent seems to be small at low latitude stations, such as Humain, Tihany, Gyergy\'o\'ujfalu, Nagycenk/Muck, Grahamstown and Sutherland. It is not known, how much of this is a result of lower detection efficiency at these stations due to local noise conditions.

Finally, we note that in some cases, there is a hint of land/sea asymmetry.
In the case of Karymshina, Figure \ref{fig:regional_westpacific}, Rothera, Halley and SANAE, Figure \ref{fig:regional_americas}, and Marion, Figure \ref{fig:regional_southernafrica}, the transmission rate seems to decrease more slowly towards the ocean than towards the continent. This may be due oceanic lightning having higher peak currents.
We would not venture into interpreting any other apparent shapes in the transmission rate maps, as they can be sensitive to individual storm events, especially where average lightning activity is otherwise low, and can show slight changes from year to year, to be investigated in our followup study. The slight differences between the transmission rates at three closely separated stations on Figure \ref{fig:regional_hu} show the limits of our method.

\begin{figure}[htb!]
    \includegraphics[width=9.5cm,height=7cm]{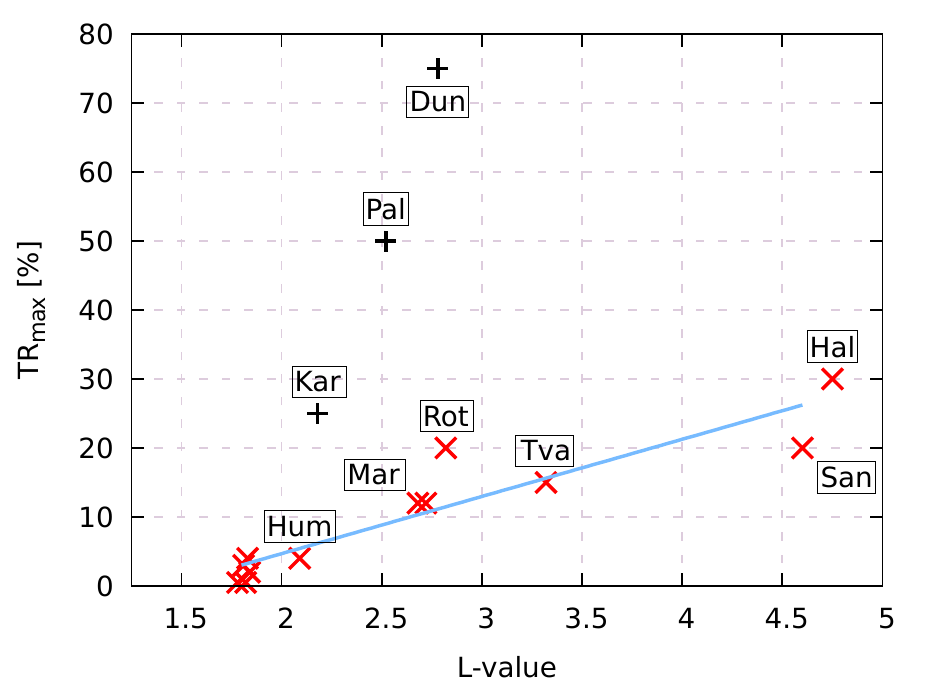}
\caption{A comparison of the obtained maximal transmission rates at each station with the L-value of the given station (see Table \ref{table:awdanetdata}). A trend of transmission rates increasing with L-values is apparent, with the exception of three points (Karymshina, Dunedin and Palmer), which we considered outliers and marked with black points. The blue line is a straight line fitted to the non-outlier points.}
    \label{fig:L-TR}
\end{figure}

In addition to listing the obtained maximal transmission rates at each station in Table \ref{table:awdanetdata}), we plotted them on Figure \ref{fig:L-TR}. A trend of transmission rates increasing with the stations' L-value is apparent, with three outlier points. At these stations, especially at Karymshina and Palmer, the VLF records show extremely  good signal quality (low background noise), which possibly contributes to the detection of low amplitude whistlers by 
the AWDANet algorithm, translating into a higher total count and therefore also a higher transmission rate. This can only be confirmed through the laborious task of visually checking large periods of raw measurement at each station.

\section{Conclusion}
We present a method to identify the general location of lightning strokes that excite detectable whistlers at ground-based receivers.
Our method also maps the geographic distribution of the lightning to whistler transmission ratio.
Our method is very general and can be applied to any dataset consisting of a whistler time series at a fixed location and a lightning database listing lightning times and locations. Our method produces results for even low number of whistlers (e.g. Eskdalemuir with $\sim 10^4$ traces, see Figure \ref{fig:regional_europe}). We applied this procedure to whistler time series recorded at fifteen ground stations over twelve years.
Our results confirm that at all of the fifteen locations the highest probability of lightning to produce a whistler detectable on the ground in the conjugate hemisphere is when the lightning is located at the geomagnetic conjugate point to the whistler observation station.
This probability decreases with increasing distance from the conjugate point.
In some cases, there is source lightning present over 3000~km from the conjugate point.

Our results are most consistent with the theory that whistlers detected on the ground propagated in a ducted mode through the plasmasphere and thus the whistler producing lightning strokes must cluster around the footprint of the ducts in the other hemisphere.
This finding has implications for the importance of ground based VLF sources, whether natural or manmade, on the loss of radiation belt electrons. It also helps clarify the application of AWDANet observations to plasmaspheric monitoring, as the AWDANet reported whistler is likely to have passed through the plasmasphere on a duct that is rather local to the AWDANet station. 

Our method has a significant potential to derive subsequent results. Having established the location of the source lightning, we can investigate how the transmission rates vary and what are they influenced by.
It is possible to look at the variation of transmission rates as a function of time of day, season, ionospheric parameters and geomagnetic activity.
Notably, our whistler database now covers more than one solar cycle, another time variable that may affect whistler transmission.
It can also be of interest to see whether the transmission rate depends on the parameters of the lightning, e. g. peak current, cloud-to-ground versus intracloud lightning, etc. Similarly, we can look at transmission rates as a function of the parameters of the whistlers, such as whistler amplitude, propagation path L-value and plasmaspheric density obtained from inverted whistler traces. These questions are outside of the scope of this paper, but we intend to investigate these ideas and present them in a follow-up study.

\acknowledgments

The authors wish to thank the World Wide Lightning Location Network (http://wwlln.net), a collaboration among over 50 universities and institutions, for providing the lightning location data used in this paper.
The research leading to these results received funding from the National Research, Development and Innovation Office of Hungary under grant agreements NN116408 and NN116446 and was supported by the ÚNKP-18-3 New National Excellence Program of the Ministry of Human Capacities of Hungary.
We acknowledge NIIF for awarding us access to supercomputing resources based in Hungary at Debrecen.

\end{document}